\documentclass{article}

\usepackage{arxiv}

\usepackage[utf8]{inputenc} 
\usepackage[T1]{fontenc}    
\usepackage{hyperref}       
\usepackage{url}            
\usepackage{booktabs}       
\usepackage{amsfonts}       
\usepackage{nicefrac}       
\usepackage{microtype}      
\usepackage{lipsum}		
\usepackage{graphicx}
\usepackage{natbib}
\usepackage{doi}

\title{Shallow Au implantation into silicon-on-insulator slot ring resonator waveguide devices}

\author{ {\includegraphics[scale=0.06]{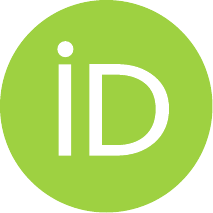}\hspace{1mm}Quan-Shan Liu}\\
	Department of Electrical and Electronic Engineering\\
	Photon Science Institute, University of Manchester\\
	Oxford Road, Manchester, M13 9PL, UK \\
	\texttt{quan-shan.liu@manchester.ac.uk} \\
	\And
	{\includegraphics[scale=0.06]{orcid.pdf}\hspace{1mm}Maddison Coke} \\
	Department of Electrical and Electronic Engineering\\
	Photon Science Institute, University of Manchester\\
	Oxford Road, Manchester, M13 9PL, UK \\
	\texttt{maddison.coke@manchester.ac.uk} \\
	\And
	{\includegraphics[scale=0.06]{orcid.pdf}\hspace{1mm}Alexander Lincoln} \\
	Faculty of Science and Engineering\\
	Electron Microscopy Centre, University of Manchester\\
	Oxford Road, Manchester, M13 9PL, UK \\
	\texttt{alexander.lincoln@manchester.ac.uk} \\
	\And
	{\includegraphics[scale=0.06]{orcid.pdf}\hspace{1mm}William Wren} \\
	Department of Electrical and Electronic Engineering\\
	Photon Science Institute, University of Manchester\\
	Oxford Road, Manchester, M13 9PL, UK \\
	\texttt{william.wren@manchester.ac.uk} \\
	\And
	{\includegraphics[scale=0.06]{orcid.pdf}\hspace{1mm}Tim Echtermeyer} \\
	Department of Electrical and Electronic Engineering\\
	Photon Science Institute, University of Manchester\\
	Oxford Road, Manchester, M13 9PL, UK \\
	\texttt{tim.echtermeyer@manchester.ac.uk} \\
	\And
	{\includegraphics[scale=0.06]{orcid.pdf}\hspace{1mm}Iain Crowe} \\
	Department of Electrical and Electronic Engineering\\
	Photon Science Institute, University of Manchester\\
	Oxford Road, Manchester, M13 9PL, UK \\
	\texttt{iain.crowe@manchester.ac.uk} \\
	\And
	{\includegraphics[scale=0.06]{orcid.pdf}\hspace{1mm}Richard J. Curry} \thanks{The corresponding author.} \\
	Department of Electrical and Electronic Engineering\\
	Photon Science Institute, University of Manchester\\
	Oxford Road, Manchester, M13 9PL, UK \\
	\texttt{richard.curry@manchester.ac.uk} \\
}

\date{}

\hypersetup{
pdftitle={A template for the arxiv style},
pdfsubject={q-bio.NC, q-bio.QM},
pdfauthor={David S.~Hippocampus, Elias D.~Striatum},
pdfkeywords={First keyword, Second keyword, More},
}

\begin{document}
\maketitle

\begin{abstract}
	The optical transmission spectra of a series of micro-ring resonators (MRRs) are studied following the implantation of gold (Au) ions and subsequent thermal annealing, at temperatures between 500 °C and 700 °C. Whilst we find that this process leads to the ready formation of Au nanoparticles (NPs) on the MRR surface, the cavity optical properties; Q-factor and extinction ratio (ER) are severely degraded, for annealing between 500 °C and 600 °C, but recover again for annealing at 650 °C. For an equivalent (control) MRR, which received no Au implantation, thermal annealing alone was also found to degrade the cavity performance.
\end{abstract}


\section{Introduction}
Silicon-based photonic devices have been a focus of intense research over recent decades, due to their compatibility with the prevailing complementary metal-oxide-semiconductor (CMOS) fabrication technology\citep{ref:siew2021,ref:xiang2021,ref:lim2013,ref:baehr2012,ref:soref2006}. Compared with other silicon photonics platforms such as silicon nitride\citep{ref:hu2017silicon}, silicon-on-sapphire\citep{ref:baehr2010silicon}, suspended silicon\citep{ref:cheng2012mid}, and chalcogenides-on-silicon\citep{ref:ma2013low}, silicon-on-insulator (SOI) is the most mature and widely offered by a number of foundries\citep{ref:siew2021}. As a result, researchers have developed a variety of SOI-based photonic building blocks\citep{maharjan2021non,poudel2023spectrometer}, among which microring resonators (MRRs) are well characterised and have been applied as functional sensors\citep{kazanskiy2023review,yebo2009silicon,kim2016cascaded,zhang2016gas,talebifard2017optimized,lo2017photonic,rodriguez2015porous,kim2010silicon,de2007silicon}, including via 2D materials integration\citep{hussein2017raman,crowe2014determination,leo2020graphene}, and more recently for the realization of novel, non-Hermitian photonics-based Parity-Time (P-T) symmetry structures\citep{li2024n,li2024modeling}. Furthermore, the slot waveguide geometry has been introduced into MRRs for achieving enhanced light-matter interaction and thus improving sensing performance\citep{robinson2008chip,liu2013highly,barrios2008label,claes2009label}.

The utilization of ion implantation techniques opens additional possibilities for the modification of material functionality\citep{namba1975ion}. This technique can be readily incorporated into CMOS processing with minimum impact on device fabrication\citep{jacobson1995high}. Through the introduction of selected atomic species with well-defined doses and selected depths, implanted matrices may display novel properties. In particular, gold has been widely implanted into a range of targets including silicon wafers\citep{mailoa2014room,hashim2013coplanar}, sapphire\citep{stepanov2005nonlinear}, silicate glasses\citep{malinsky2009implantation}, Nd: YAG crystals\citep{nie2018plasmonic}, TiO$_{2}$ crystal\citep{ye2022embedded}, LiTaO$_{3}$ crystals\citep{pang2021q}, silica\citep{magruder1993optical,torres2015collective}, and polymethyl methacrylate resist\citep{adshead2023high}. It has been observed that with suitable post-implantation annealing Au nanoparticles (NPs) may be formed displaying plasmonic properties \citep{nie2018plasmonic,wang2024effects,ribeiro2023plasmonic,hsieh2016optical,ramaswamy2005synthesis}. However, gold implantation directly into MRRs has remained unexplored, though some attention has been paid to other dopants including oxygen\citep{waldow200825ps}, germanium\citep{reed2017trimming,milosevic2018ion}, silicon\citep{ackert2011defect}, and boron\citep{ackert2011defect,hagan2019post}.

In this paper we report experimental studies of gold implantation into SOI-based slot MRRs with the aim of Au NP formation and modification of the MRR optical response. The transmission spectra of MRR devices at each stage of processing (pre/post-implant and post-annealing) are measured to obtain extinction ratios (ER) and quality factors (QF) to assess the impact of the processing. Though in this study we found no evidence of any enhanced plasmonic effect associated with the coupling of Au NPs to the MRRs, the impact of the processing on the MRR optical response is elucidated.

\section{Sample details and methods}
All the waveguide devices studied were rib-type and designed for single-mode operation around 1550 nm. Based on a commercial 220 nm SOI platform, the rib waveguides were on-chip fabricated by deep ultraviolet lithography, with a 130 nm rib height and 90 nm slab thickness on top of a 2 \textmu m buried oxide layer. The chip was coated with silica cladding of \textasciitilde 1.5 \textmu m thickness for device protection.

\subsection{Device structure and ion implantation}
\begin{figure}[ht!]
    \centering\includegraphics[width=16cm]{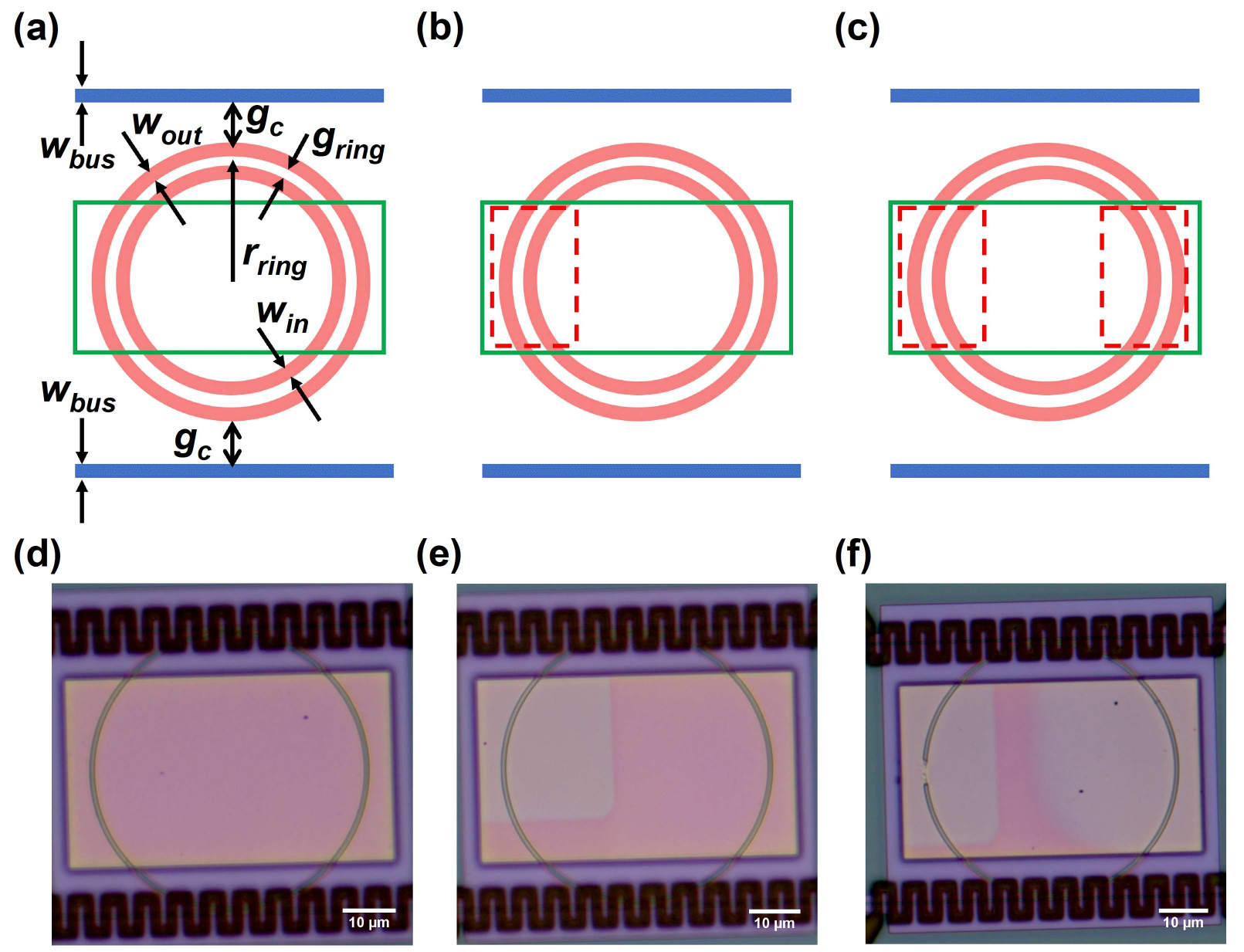}
    \caption{(a) to (c) Plan-view not to scale schematic illustration of three add-drop slot MRR devices with the device design parameters defined (values provided in the text). The blue lines represent the bus waveguides adjacent to the MRRs shown in pink. The etched cladding region is shown by the green box with the red dashed boxes indicating the regions implanted with Au ions. (d) to (f) Optical microscope images of each device post-implantation corresponding to (a) to (c). The implanted area is observed as the colourless region contrasting with the pink-tinged unimplanted regions.}
    \label{Fig1}
    \end{figure}

The slot MRRs used in this study can be classified into two configurations consisting of (i) ‘single-input-through' and (ii) ‘add-drop' designs. Figures \ref{Fig1}a to \ref{Fig1}c show schematic plan-views of three add-drop slot MRRs, with the structural parameters defined as follows: inner ring waveguide width $w_{in}$ = 290 nm; outer ring waveguide width $w_{out}$ = 250 nm; nearby bus waveguide width $w_{bus}$ = 320 nm; ring-to-ring gap $g_{ring}$ = 200 nm; ring-to-bus gap $g_{c}$ = 200 nm, 250  nm, and 300 nm for Figures \ref{Fig1}a to \ref{Fig1}c respectively; and radius to slot centre $r_{ring}$ = 25 \textmu m. To allow ion implantation into the MRRs a series of 64 \textmu m $\times$ 35 \textmu m etched windows were opened through the silica capping layer, shown by the green box in Figure \ref{Fig1}a-c and visible in the optical images in  Figure \ref{Fig1}d-f. The slot MRR device in Figure \ref{Fig1}a remained unimplanted as a reference (referred to as device A); the two devices in Figure \ref{Fig1}b and \ref{Fig1}c were implanted with 25 keV Au+ ions into a region covering \textasciitilde18$\%$ (Figure \ref{Fig1}b, device B) and \textasciitilde46$\%$ (Figure \ref{Fig1}c, device C) of the total MRR with a dose of 5E14 ions/cm$^{2}$. Device C appears to have a discontinuity in its ring that was not noticed prior to characterization. Surprisingly, resonances are still measurable for this "incomplete-ring" structure, with wider peak widths compared to those of devices A and B (to be shown in later figure). 

A second set of single bus MRR devices were likewise implanted for study, again one of these single-input slot MRR devices remained unimplanted providing a reference (referred to as device D). The two other devices (referred to as devices E and F respectively) were implanted with 25 keV Au+ ions into a region covering \textasciitilde22$\%$ and \textasciitilde49$\%$ of the total MRR respectively, with a higher dose of 1E15 ions/cm$^{2}$. The percentages of coverage are calculated based on the implantation areas presented in the optical figures, and the details of implantation of all the devices mentioned are summarized in Table \ref{device details}.

    \begin{table}[!ht]
        \centering
        \caption{Summary of implantation details for each device.}
        \begin{tabular}{ccccc}
        \hline
            Device & Implantation dose (ions/cm$^{2}$) & Comments  \\ \hline
            A & - & add-drop; reference device  \\ 
            B & 5E14 & add-drop; covering \textasciitilde18$\%$ of the total MMR \\
            C & 5E14 & add-drop; covering \textasciitilde46$\%$ of the total MMR \\
            D & - & single-input; reference device  \\
            E & 1E15 & single-input; covering \textasciitilde22$\%$ of the total MMR \\
            F & 1E15 & single-input; covering \textasciitilde49$\%$ of the total MMR \\ \hline
        \end{tabular}
        \label{device details}
    \end{table}

The ion implantation was performed using the platform for nanoscale advanced materials engineering (P-NAME) Facility (Ionoptika, Q-One) at the University of Manchester\citep{adshead2023high}. Inspection of the optical microscope images exhibits a clear contrast between implanted and unimplanted areas (clear and pink), illustrated in Figure \ref{Fig1}d to \ref{Fig1}f for devices A to C respectively. Post-implantation thermal annealing is typically required for both repairing ion damage and electronic activation of implanted dopants\citep{mayer1970eriksson}. The selection of optimal annealing conditions is highly empirical and varies depending on the specific implanted species and the host material. In this case the temperature and duration must be sufficient to stimulate the migration of the implanted Au ions to form NPs whilst not resulting in damage to the waveguides. In order to determine a starting point for annealing whilst preserving MRRs, 25 kV Au+ was first implanted, at a dose of 5E15 ions/cm$^{2}$, into a simple on-chip linear waveguide along a 500 \textmu m length. Annealing was then performed under a nitrogen atmosphere at 500 °C. 

To ascertain that these annealing conditions are suitable for the formation of Au NPs SEM imaging of the implanted waveguides was undertaken (Supplementary Information, Figures \ref{FigS1}a and \ref{FigS1}b). Close inspection revealed the formation of Au NPs with a diameter of \textasciitilde10 nm on the sample surface. Energy dispersive x-ray spectroscopy (EDS) was used to compare waveguides with and without Au implantation with an enhancement of the Au x-ray signal observed (Supplementary Information, Figure \ref{FigS1}c) in the implanted area. These results indicated that a 500 °C anneal is a suitable temperature to initiate Au NP formation however, this alone may not be sufficient to repair implantation damage. To assess this the transmission spectra of the linear waveguide were measured pre- and post-implantation and again post-anneal. It is found that implantation leads to sufficient damage to prevent waveguiding which is not recovered by a 500 °C anneal.

To study the MRR devices A to F (Table \ref{device details}) a series of thermal anneals were performed under a nitrogen atmosphere each lasting for 2 minutes at temperatures between 500 °C to 700 °C at 50 °C increments, summarised in Table \ref{stage definition}. The two unimplanted reference devices (A and D) were also subject to the same thermal annealing.

    \begin{table}[!ht]
        \centering
        \caption{Summary of annealing cycles undertaken.}
        \begin{tabular}{ccccc}
        \hline
            Stage & Definition & Comments  \\ \hline
            1 & Pre-implant & -  \\ 
            2 & Post-implant & Devices A and D remained unimplanted \\
            3 & Post-anneal under 500 °C & N$_2$ atmosphere for 2 minutes \\
            4 & Post-anneal under 550 °C & N$_2$ atmosphere for 2 minutes \\
            5 & Post-anneal under 600 °C & N$_2$ atmosphere for 2 minutes \\
            6 & Post-anneal under 650 °C & N$_2$ atmosphere for 2 minutes \\
            7 & Post-anneal under 700 °C & N$_2$ atmosphere for 2 minutes \\ \hline
        \end{tabular}
        \label{stage definition}
    \end{table}

\begin{figure}[ht!]
    \centering\includegraphics[width=16cm]{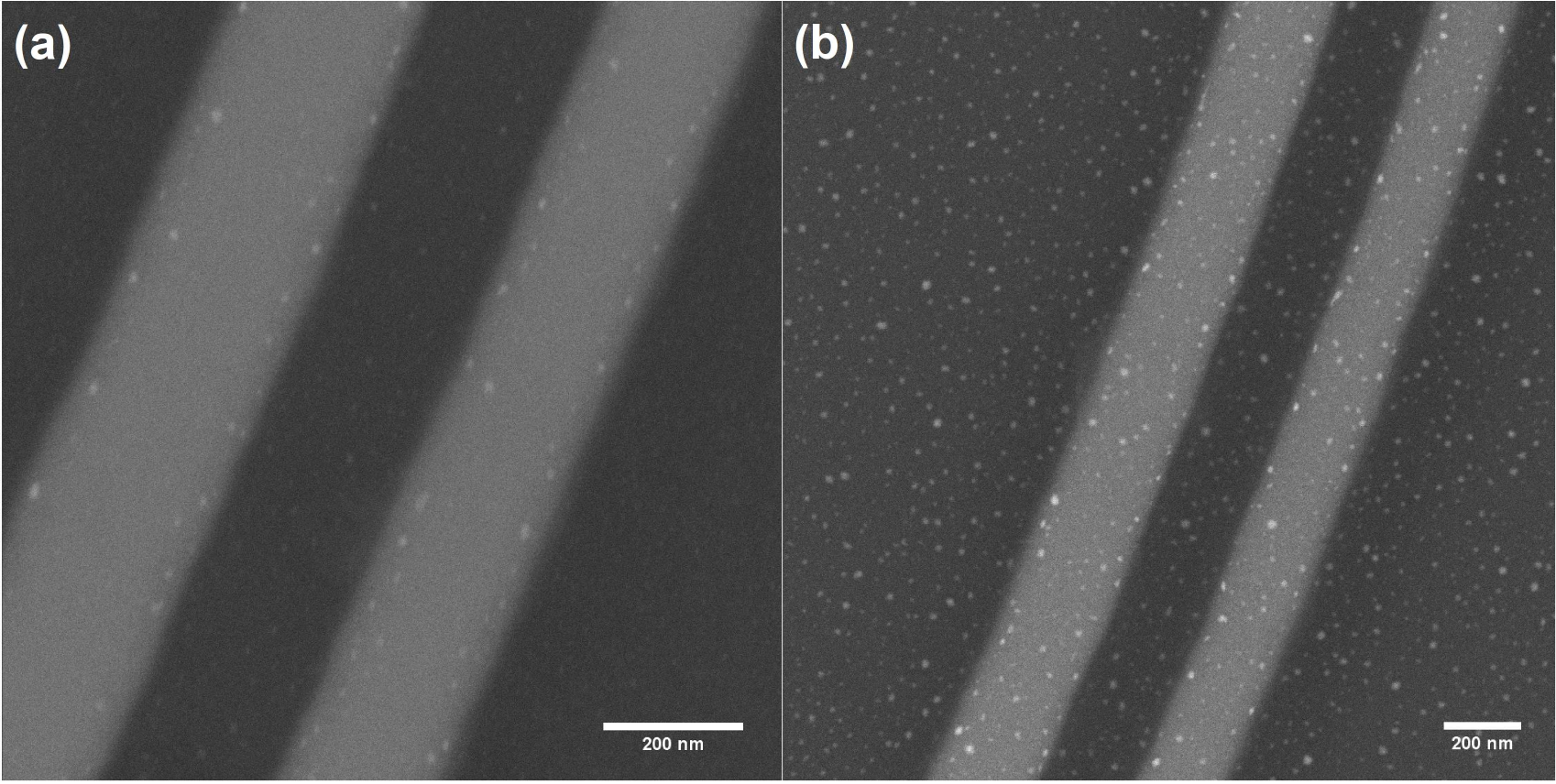}
    \caption{(a) Back scattered electron image for the implanted ring waveguide of device C. (b) back scattered electron image for the implanted ring waveguide of device F. Both devices have gone through a number of 2 min thermal anneals between 500 °C and 700 °C at 50 °C intervals (stages 1 to 7 in Table \ref{stage definition}).}
    \label{Fig2}
    \end{figure}

Following each stage MRR characterisation was performed, described below, and post stage 7 SEM images of devices C and F were obtained (Figure \ref{Fig2}). Within each image sporadic Au NP decoration can be observed, increasing in density with Au dose (5E14 and 1E15 ions/cm$^{2}$ for devices C and F respectively). Alongside this, EDS analysis was performed on the implanted and unimplanted areas of devices B and E (Supplementary Information, Figures \ref{FigS2}a and \ref{FigS2}b) showing an increase of the Au signal in the implanted regions. 

\section{Experimental set-up}
To quantify the impact of the Au implantation and thermal annealing on the MMR devices the transmission of each device's bus waveguide was measured at each stage of processing. For the add-drop slot MRRs, the same bus waveguide was consistently selected for characterization after each stage. A Thorlabs 1550 nm fibre-coupled benchtop superluminescent diode coupled to a single-mode cleaved-end fibre was used to launch light into the waveguide via the predefined grating couplers. Transmitted light was similarly collected using a multi-mode fibre and inserted into an Anritsu optical spectrum analyser. All measurements were taken at room temperature and over the 1450 nm to 1650 nm spectral range.

The measured input and output wavelength dependent power ($P_{in}$ and $P_{out}$ respectively) is used to obtain the transmission loss of waveguides (including input and output grating couplers), as:

    \begin{equation}
        Loss = 10\times log_{10}(P_{out}/P_{in})
        \label{equ:loss}
    \end{equation}

Figure \ref{Fig3}a shows the $P_{in}$ and $P_{out}$ of the unimplanted reference device A during one measurement. Each spectrum was collected three times, performing a fresh fibre alignment each time, to obtain the mean value and standard deviation.

\section{Data processing and analysis}
\begin{figure}[ht!]
    \centering\includegraphics[width=16cm]{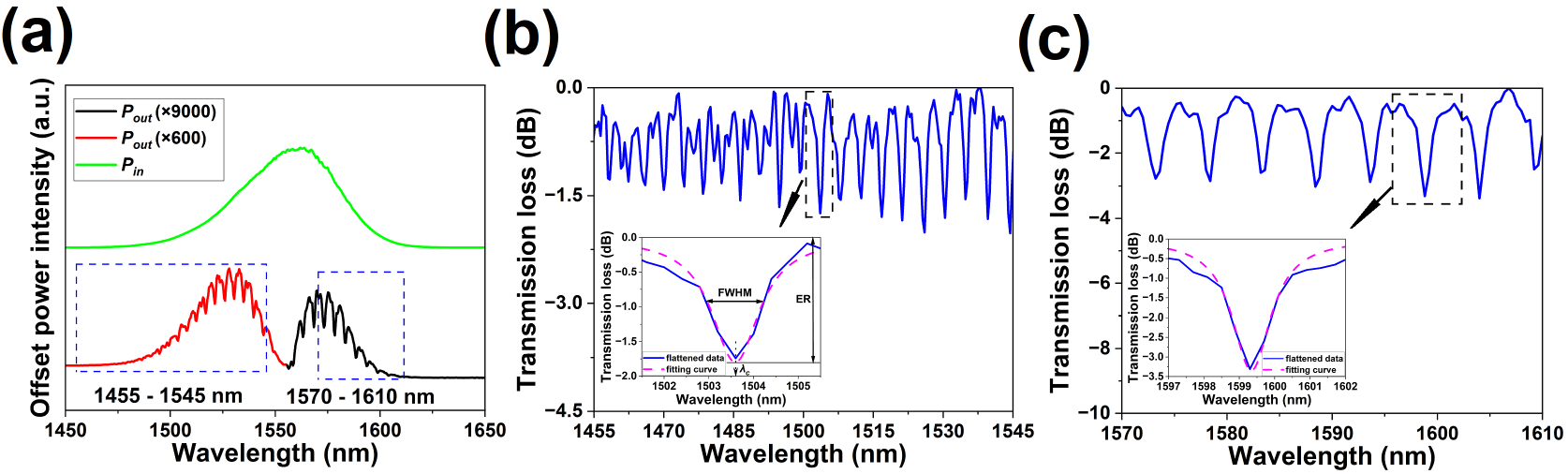}
    \caption{(a) The input and output power intensity of reference device A as received (vertically offset for clarity). The two selected regions over which background subtraction and analysis was performed are shown enclosed by blue dashed boxes. (b) and (c) Demonstration of spectrum flattening and resonance peak fitting for the two selected regions displayed in (a).}
    \label{Fig3}
    \end{figure}

In Figure \ref{Fig3}a, the grating coupler bandwidth is not in perfect alignment with the light source. Nevertheless, resonant patterns are observed both below and above 1550 nm. As such, two selected regions (1455 to 1545 nm, and 1570 to 1610 nm) of each measured spectra were chosen for data processing and analysis, each displaying multiple resonance peaks. The processing removed the background response within the selected regions using subtraction of a polynomial fit resulting in a flattening of the spectrum, Figure \ref{Fig3}b and \ref{Fig3}c. This step does not change any resonance properties and is helpful for resonance comparison among different spectra. The ER and QF were derived from Lorentzian fitting of the resonance peaks based on the least square method using MATLAB codes modified from that provided in reference \citep{alsalman2021}. 

Figures \ref{Fig3}b and \ref{Fig3}c show the flattened data (blue line) and the fitted Lorentzian curves to one resonance (magenta dashed line) for the two marked regions in Figure \ref{Fig3}a. Relevant parameters including one peak's ER, full-width half maximum (FWHM), and central resonant wavelength ($\lambda$$_c$) are illustrated in Figure \ref{Fig3}b inset. Separate spectrum flattening and curve fitting is completed for the two selected wavelength regions. To ensure consistent convergence of peak fitting, only resonances exceeding 70\% of the deepest resonance peak in each region are selected for fitting.

\section*{Results and discussion}
\subsection{Transmission spectrum analysis}
\begin{figure}[ht!]
    \centering\includegraphics[width=16cm]{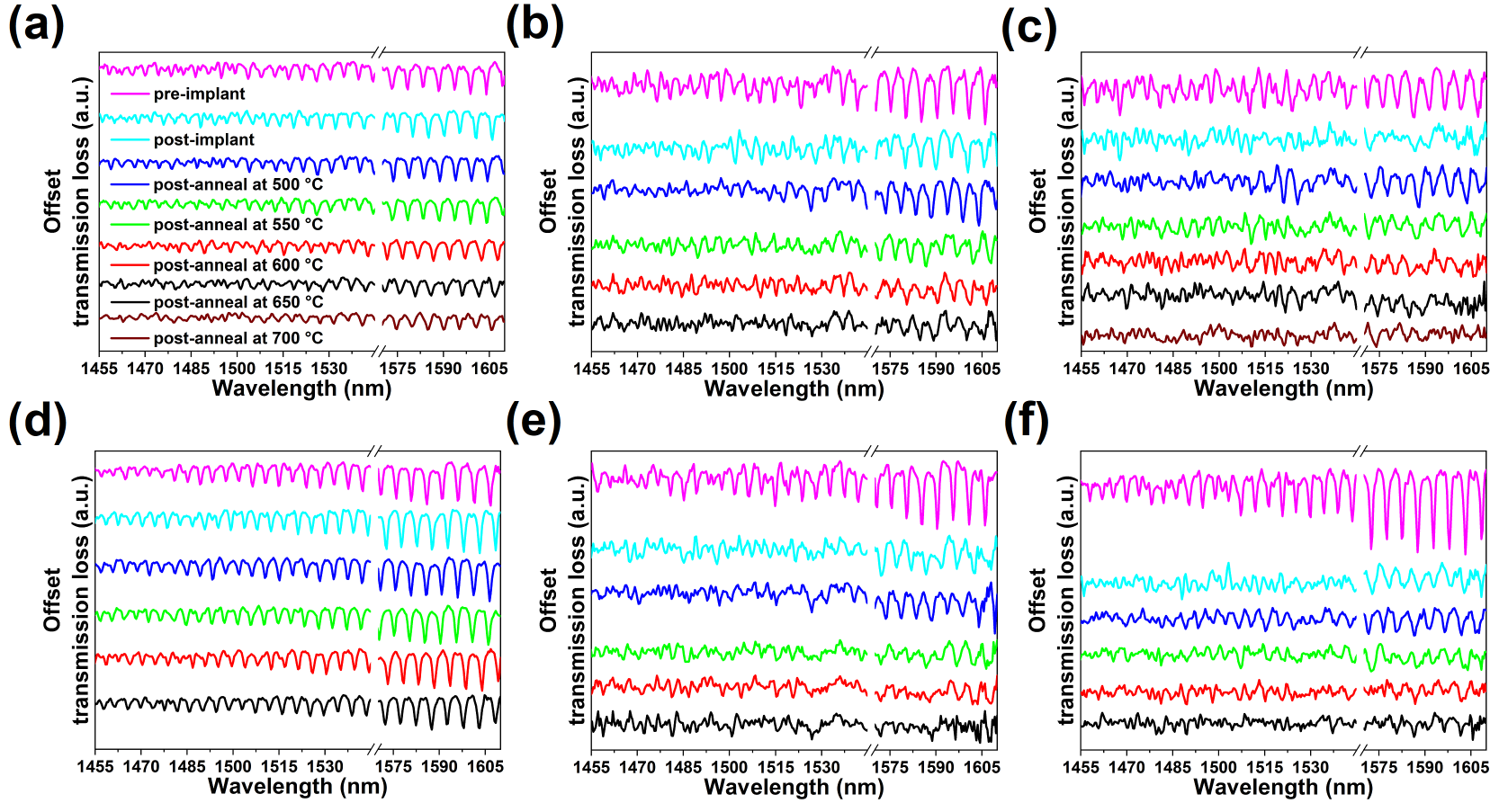}
    \caption{(a) to (f) Flattened transmission spectra of devices A to F respectively. Within each panel the stages progress from top (stage 1) to bottom as indicated in (a) and are vertically offset for clarity.}
    \label{Fig4}
    \end{figure}

Figures \ref{Fig4}a to \ref{Fig4}f show the flattened resonance peaks of devices A to F (Table \ref{device details}) respectively at all stages of processing (see Table \ref{stage definition} for stage definition). The transmission of devices B, D, E, and F following stage 7 (700 °C annealing) are not included as these devices exhibited no transmission. This clearly indicates device failure resulting from thermally induced damage caused by thermal expansion coefficient mismatch between the Si and SiO$_2$. Damage can be visibly observed by comparing optical microscope images of the devices following stage 1, 6, and 7 (Supplementary Information, Figure \ref{FigS3}).

Following implantation (stage 2) but prior to any annealing, all implanted devices experience a reduced coupling to the MRR with devices C and F (higher doses) reducing more than devices B and E (lower doses). The QFs of both of the unimplanted reference devices are also found to be slightly different which might be attributed to variation in laboratory conditions (e.g. humidity) between measurements. Following 500 °C annealing it is observed that the MRR resonance peaks have recovered somewhat, indicating a reduction in ion-induced damage. Additionally, at this temperature Au ions have diffused out of the waveguides to form Au NPs on the device surface as shown in Figure \ref{Fig2}. Annealing at higher temperature (stages 3 to 7) results in a general degradation of the MRR ER. For the reference unimplanted devices, a slight degradation in MRR QF can be seen as annealing progresses. To quantify the above outlined impact of annealing on the devices Lorentz fitting of the resonance peaks was performed.

\subsection{Extinction ratio analysis}
\begin{figure}[ht!]
    \centering\includegraphics[width=16cm]{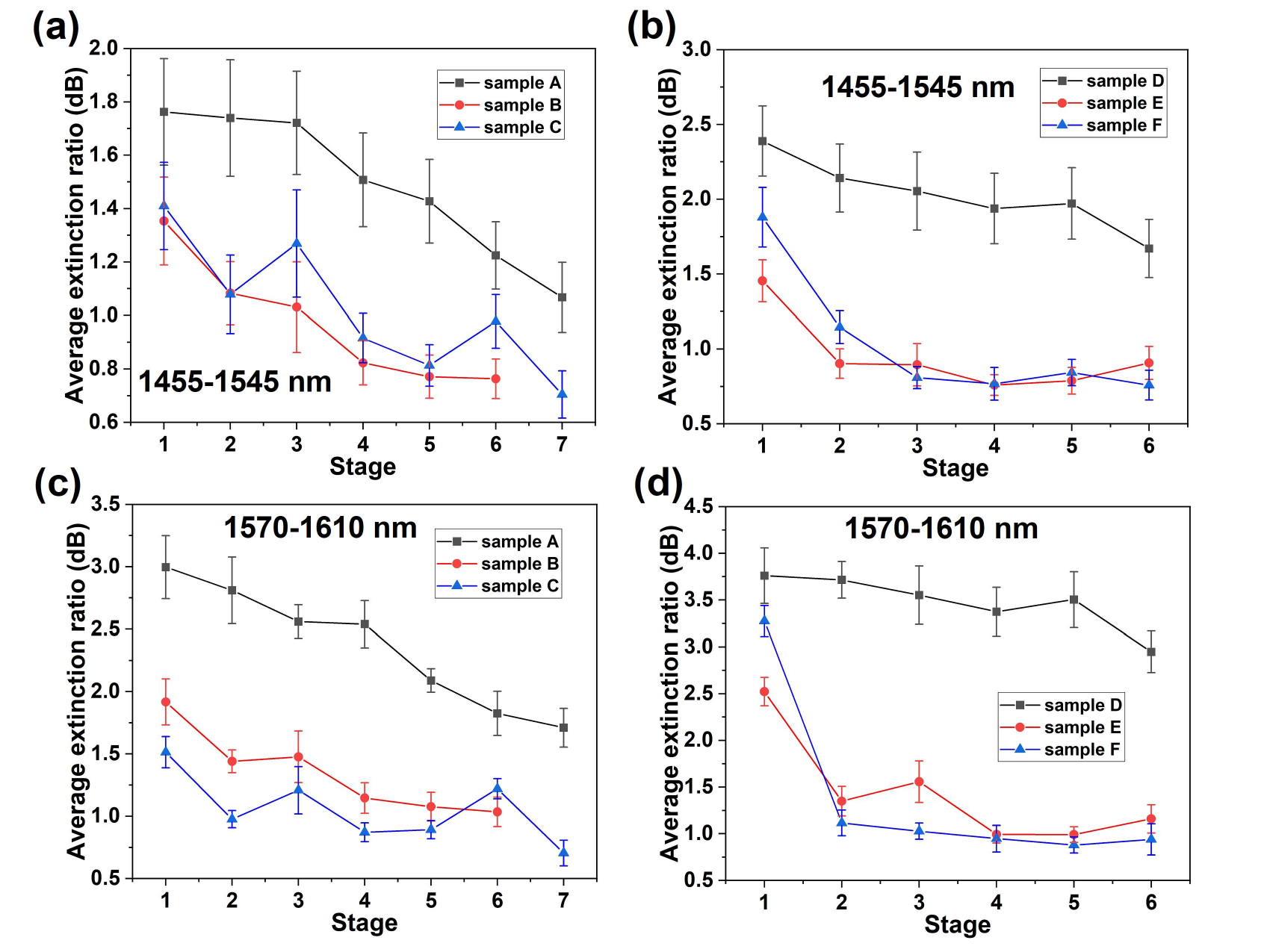}
    \caption{The average ERs of (a) devices A to C in the 1455 to 1545 nm region, (b) devices D to F in the 1455 to 1545 nm region, (c) devices A to C in the 1570 to 1610 nm region and (d) devices D to F in the 1570 to 1610 nm region following each experimental stage (Table \ref{stage definition}).}
    \label{Fig5}
    \end{figure}

 To extract ERs from the resonance peaks displayed in Figure \ref{Fig4}, Lorentz curve fitting was performed as described above. Figure \ref{Fig5} displays the average ER within each wavelength region analysed for each of the different experimental stages performed, with error bars representing the standard deviation.

As shown in Figure \ref{Fig5}a and \ref{Fig5}b, the average ERs measured across the 1455 nm to 1545 nm region of devices B, C, E, and F reduce by ~20$\%$, 24$\%$, 38$\%$, and 39$\%$ following implantation. A similar trend is displayed within the 1570 nm to 1610 nm region, Figure \ref{Fig5}c and \ref{Fig5}d, but with greater reduction of the ERs by ~25$\%$, 36$\%$, 47$\%$, and 66$\%$ for devices B, C, E, and F respectively. The reduction in ER increases with implantation dose and area, strongly linking it to ion implantation induced damage within the MRRs. Annealing at 500 °C does not repair this damage as mentioned above for the test linear waveguide, however Au NPs are formed on the MRR surface. Inspection of the bus waveguide transmission spectra before and after implantation, and following 500 °C annealing, did not reveal any obvious spectral difference in response. Plasmonic coupling to the Au NPs surface plasmon modes is ruled out as the wavelengths used are not in resonance with the Au surface plasmon energy (~515 nm to 570 nm for 10 nm to 100 nm diameter NPs respectively). 

To see if higher annealing temperatures can be effective in recovering performance the further sequential anneals were performed up to 700 °C. Following each of the subsequent annealing stages the ERs of both of the unimplanted reference devices exhibit a steady reduction. Those of the implanted devices B and C likewise continue to decline with post implantation annealing. In contrast the ERs of devices E and F appear to remain similar to their post implantation values. High-temperature annealing is expected to lead to some structural deformation due to the thermal expansion coefficient mismatch within the chip resulting in increased loss. However, annealing also recovers ion-induced damage and can lead to surface smoothing which should reduce loss. These effects are in competition and would appear to be balanced for devices E and F.

\subsection{Quality factor analysis}

\begin{figure}[ht!]
    \centering\includegraphics[width=16cm]{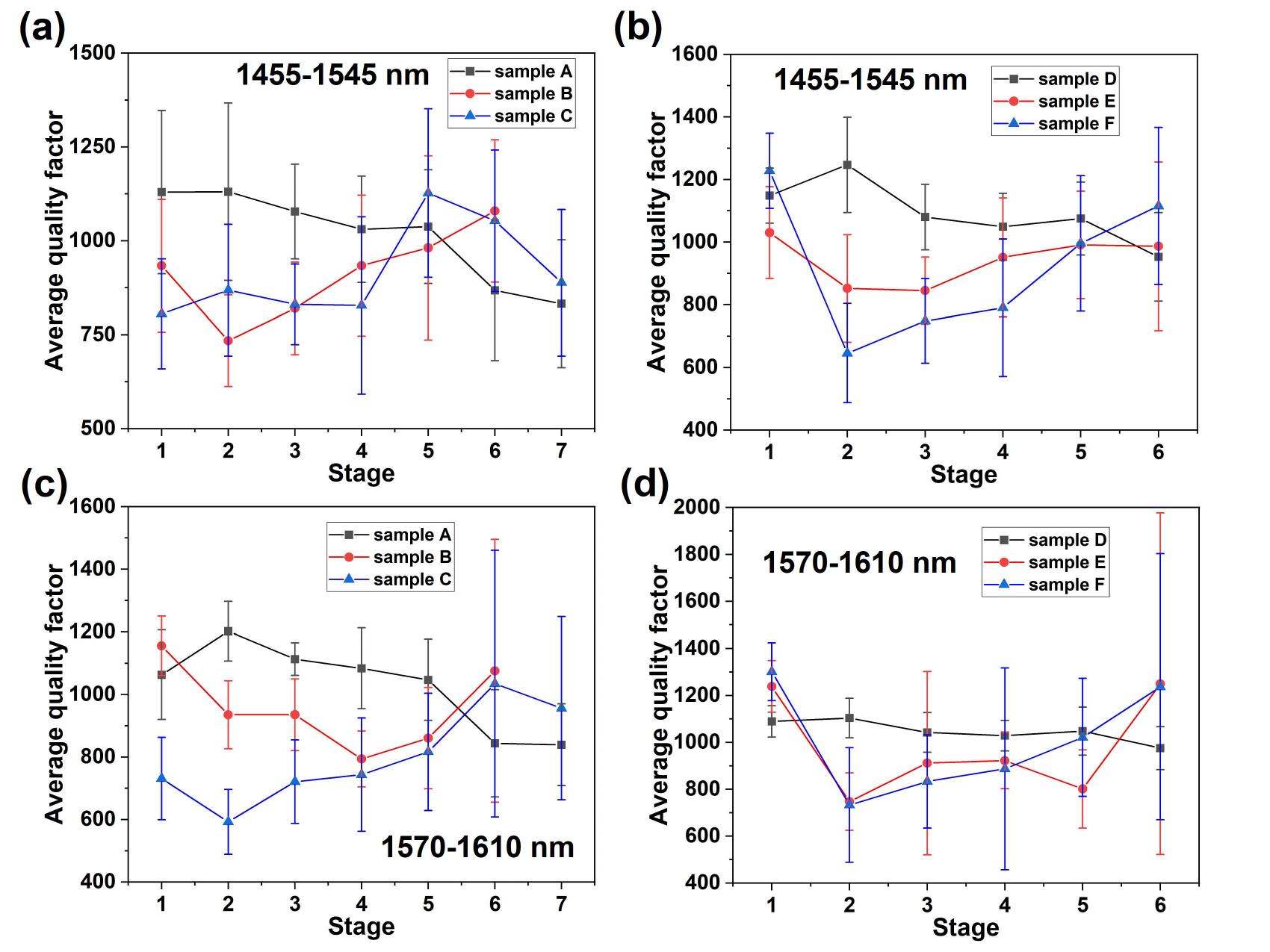}
    \caption{The average QFs of (a) devices A to C in the 1455 to 1545 nm region, (b) devices D to F in the 1455 to 1545 nm region, (c) devices A to C in the 1570 to 1610 nm region and (d) devices D to F in the 1570 to 1610 nm region following each experimental stage (Table \ref{stage definition}).}
    \label{Fig6}
    \end{figure}

The quality factor (QF) is defined as the ratio of the central resonant wavelength to the full-width half maximum of the resonance. High QFs are usually desired to realise ultra-sensitive photonic devices. QFs were extracted from the Lorenzian fits to the resonance peaks shown in Figure \ref{Fig4}. Figure \ref{Fig6} presents the average QFs of each device at each experimental stage, with the error bars indicating the standard deviations.

The variation of the QFs with implantation and annealing is not as obvious as that of the ERs with most variation occurring within the measurement error confidence. For the unimplanted devices, Figures \ref{Fig6}a and \ref{Fig6}b suggest that the QF of reference device A undergoes only a small reduction until an obvious decrease at stage 6 (annealing at 650 °C). The QF of reference device D is barely affected by annealing. Following ion implantation, a clear reduction in QF is generally observed for all devices as shown in Figure \ref{Fig6}. There appears to be no strong relationship between the magnitude of the QF reduction and implantation dose or area. As annealing proceeds up to 650 °C, the overall outcome is for the QFs of implanted devices to return to a similar value as that measured pre-implantation. Further annealing at a higher temperature (700 °C, stage 7) does not seem to result in any further QF increase but did damage several devices. Further analysis is impeded by relatively large error bars which are possibly caused by the uneven distribution of peak widths indicated in Figure \ref{Fig4}.

\section{Conclusion}
We have explored the feasibility of forming Au NPs on the surface of SOI slot MRR waveguides through the use of FIB implantation. The presence of the Au NPs following 25 keV Au implantation at doses of 5E14 and 1E15 ions/cm$^2$ and following a 500 °C thermal anneal was verified via SEM imaging and EDS analysis. The impact of ion implantation on the waveguiding properties of the MRR systems is to reduce their cavity ER and QF, as would be expected due to ion induced damage. Thermal annealing alone, at temperatures from 500 °C to 700 °C, is also observed to result in a reduction of the waveguiding and MRR performance indicating that damage is occurring which we attribute to thermal expansion mismatch within the structures. Annealing of the Au implanted devices is found to result in a further reduction in waveguiding as measured by the ERs of the measured devices. In contrast annealing at a temperature of 650 °C was able to almost fully recover the measured QFs of implanted devices. Whilst this work demonstrated the ability to form Au NPs through direct writing of Au ions into MMR devices, the resulting optical properties are degraded by both the implantation itself and the thermal treatment. 

\section{Acknowledgements}

This work was funded by EPSRC grant EP/V001914/1, and by capital investment by the University of Manchester. Q-S.L thanks the China Scholarship Council for financial support.

\section{Author contributions statement}

Q-S.L, M.C, A.L, and W.W conducted all experimental work. Q-S.L, M.C and R.C performed data analysis. T.E, I.C and R.C supervised the work. All authors contributed to the writing of the manuscript.

\section{Competing interests}

The author(s) declare no competing interests.

\bibliographystyle{unsrtnat}
\bibliography{references}  

\clearpage
\begin{center}
    \Large
    \textbf{Supplementary information}
    \\[20pt]
    \normalsize 
\end{center}
\setcounter{figure}{0}
\renewcommand*{\thefigure}{S\arabic{figure}}

\begin{figure}[htbp]
\centering
\includegraphics[width=16cm]{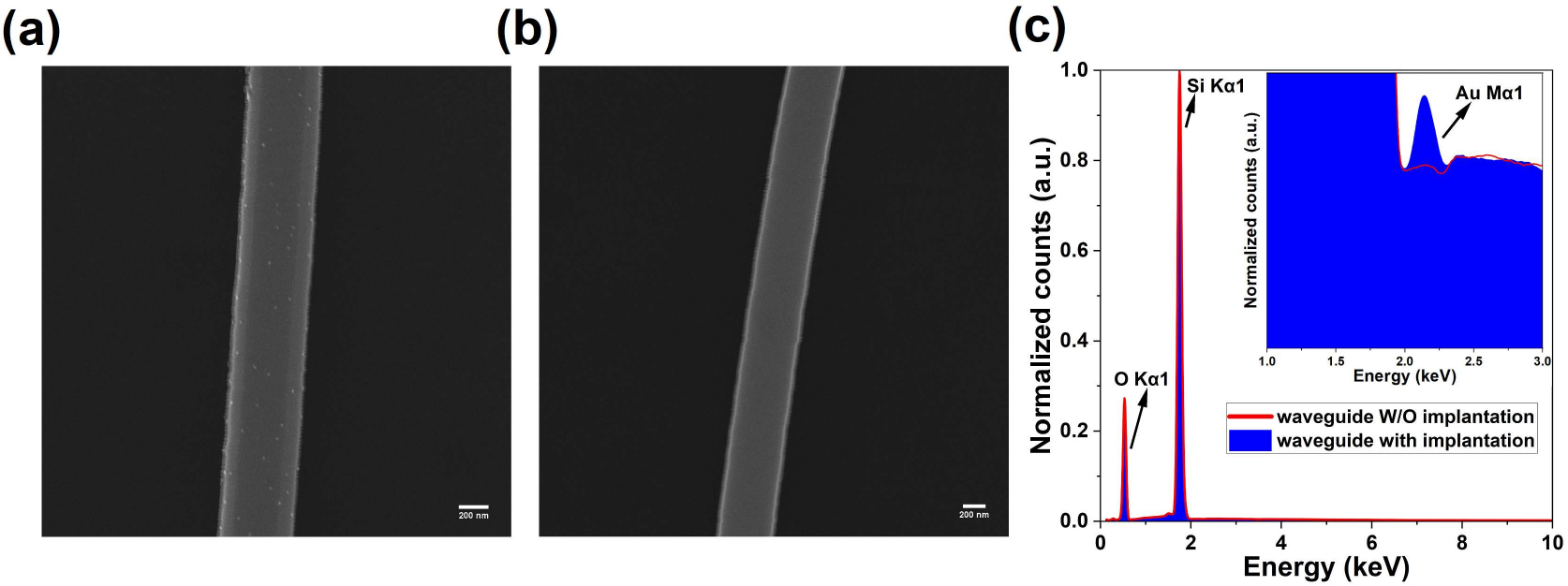}
\caption{Post-annealing (500 °C) secondary electron images of waveguides (a) following 5E15 cm$^{-2}$ gold implantation, (b) without gold implantation. (c) EDS spectra of the devices mentioned in (a) and (b) showing clear increase in Au signal in the implanted device.}
\label{FigS1}
\end{figure}

\begin{figure}[htbp]
\centering
\includegraphics[width=16cm]{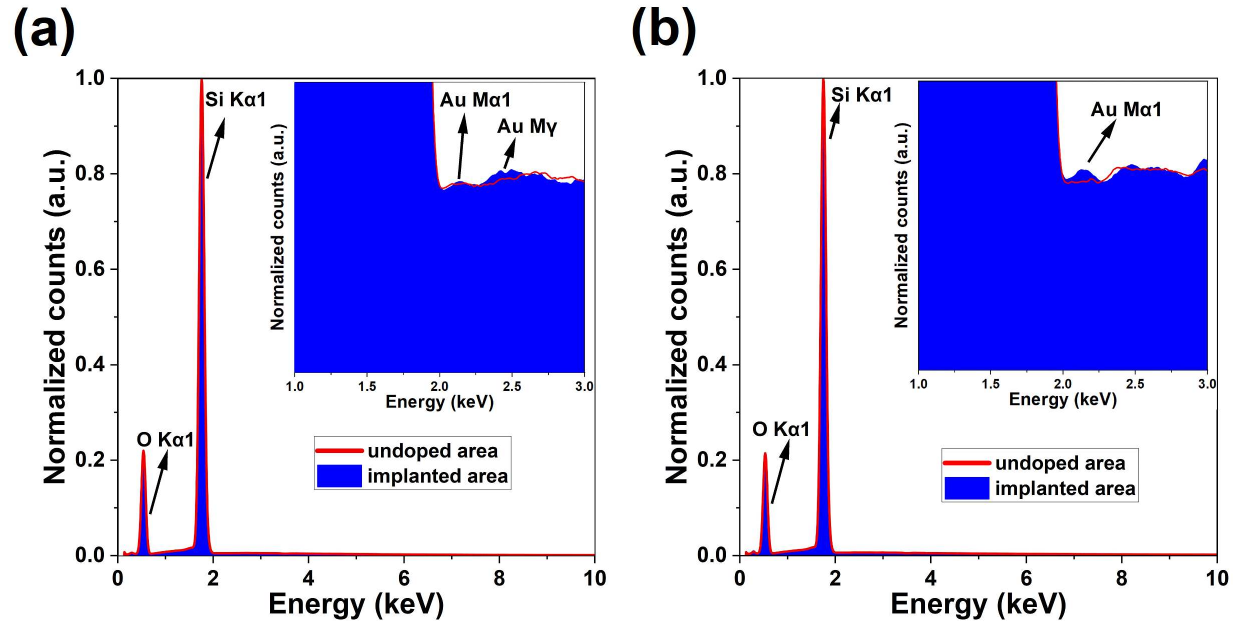}
\caption{Post-annealing EDS spectra for the unimplanted and Au implanted areas of (a) device B, (b) device E.}
\label{FigS2}
\end{figure}

\begin{figure}[htbp]
\centering
\includegraphics[width=12cm]{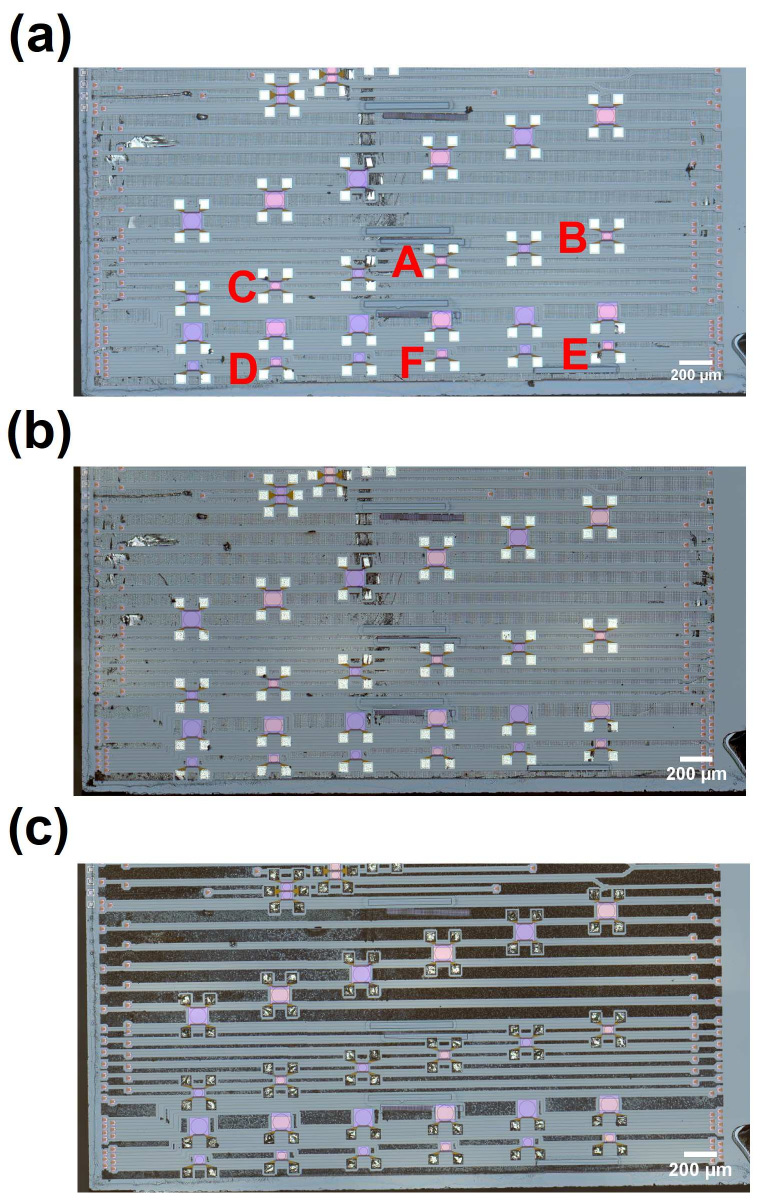}
\caption{Optical images of the photonic chip showing devices studied, labeled to the left of each MRR. (a) Image taken prior to implantation. (b) Image taken following implantation (except reference devices A and D) and a series thermal anneals from 500 °C to 650 °C. (c) Image taken following a further anneal at 700 °C. In (c) it can be observed that the chip has been thermally damaged by the contrasting dark regions with respect to (a) and (b).}
\label{FigS3}
\end{figure}






\end{document}